 \documentclass[preprints,article,accept,moreauthors,pdftex]{Definitions/mdpi}

\firstpage{1} 
\pubvolume{xx}
\issuenum{1}
\articlenumber{5}
\pubyear{2020}
\copyrightyear{2020}
\history{Received: 05 July 2020; Accepted: 01 August 2020; Published: 03 August 2020}
\Title{The~Habitability of the Galactic Bulge}

\Author{Amedeo Balbi 
 $^{1,}$*, Maryam Hami $^{1}$ and Andjelka Kova{\v c}evi{\'c} $^{2}$}

\AuthorNames{Amedeo Balbi, Maryam Hami and Andjelka Kova{\v c}evi{\'c}}

\address{%
$^{1}$ \quad Dipartimento di Fisica, Universit\`a di Roma ``Tor Vergata'', Via della Ricerca Scientifica, 00133 Roma, Italy; \\
$^{2}$ \quad Department of Astronomy, Faculty of Mathematics, University of Belgrade, Studentski trg 16, 11000~Belgrade, Serbia; }
\corres{Correspondence: balbi@roma2.infn.it}

\abstract{We present a new investigation of the habitability of the Milky Way bulge, that expands previous studies on the Galactic Habitable Zone. We discuss existing knowledge on the abundance of planets in the bulge, metallicity and the possible frequency of rocky planets, orbital stability and encounters, and the possibility of planets around the central supermassive black hole. We focus on two aspects that can present substantial differences with respect to the environment in the disk: (i)~the ionizing radiation environment, due to the presence of the central black hole and to the highest rate of supernovae explosions and (ii) the efficiency of putative lithopanspermia mechanism for the diffusion of life between stellar systems. We use analytical models of the star density in the bulge to provide estimates of the rate of catastrophic events and of the diffusion timescales for life over interstellar~distances.}


\begin{document}

\section{Introduction}
With the growing number of exoplanets discovered in the past two decades, the question of what factors make an environment conducive to life has received increasing attention. In addition to exploring the question of habitability in a planetary context, many studies have investigated it from a broader perspective, with the purpose of identifying locations in the Milky Way where the chance of finding habitable worlds is higher. 

In analogy with the habitable zone around a star~\cite{Kasting1993}, the region encompassing habitable locations in the galaxy has been termed Galactic Habitable Zone (GHZ). So far, different theoretical approaches have not converged onto a single, well-defined GHZ model. For~example, while early studies on the GHZ have identified an annular region in the Galactic disk, roughly centered around the Sun location, as the most likely location for habitable exoplanets~\cite{Gonzalez2001, Lineweaver2004}, subsequent works have deemed the entire disk almost equally habitable      \cite{Prantzos2008, Forgan2017}, while others suggested that habitability is more likely at the outskirts of the galaxy~\cite{Vukotic2016}. For~a recent review of the various approaches and results on the GHZ, and for a discussion of the divergence in the literature, we refer the reader to~\cite{Gowanlock2018}.  

So far, little attention has been devoted to the habitability of the innermost region of the galaxy. Generally, locations closer to the Galactic center have been judged unfavorably in terms of habitability. This is mainly due to the highest rate of potentially hazardous astrophysical events (e.g., nearby supernovae explosions) and to the increasing content of heavier elements (i.e., metallicity). The~latter is thought to cause an overproduction of gas giant planets and hamper the development of terrestrial planets~\cite{Lineweaver2004}. However, it has also been argued~\cite{Gowanlock2011, Morrison2015} that the highest density of star systems in the inner Galactic regions could result in an increased window of opportunity for life to emerge there, with respect to the outskirts, despite the higher risk of sterilizing events. Another concern for habitability is the presence of the supermassive black hole in the Galactic center, that could have resulted in a substantial flux of ionizing radiation during its past active phase~\cite{Balbi2017, Forbes2018, Wislocka2019}, causing increased planetary atmospheric erosion and potentially harmful effects to surface life. At the same time, the presence of such a high-energy source might have had positive effects, e.g., by increasing the number of terrestrial planets formed from the evaporation of the gas envelop of sub-Neptune planets~\cite{Chen2018}, or by stimulating prebiotic chemistry or even photosynthetic activity on otherwise poorly irradiated planets~\cite{Lingam2019a}. 

In addition to these general considerations, most previous studies on the GHZ have skipped altogether the investigation of regions within roughly 2 kpc of the Galactic center. In this { article}, we focus our attention precisely on the habitability of such innermost regions, and especially on the Galactic bulge. First, we review what is currently known about the possible presence of planets around stars in the bulge. Next, we explore the factors that have been deemed relevant in previous studies of Galactic habitability, and we investigate their importance in the bulge. We then focus on the radiation environment in the bulge, arising from the presence of the central supermassive black hole and from supernovae explosions. Finally, we investigate the efficiency of the lithopanspermia mechanism as a possible way to exchange biological material between stellar systems. 


\section{Planets in the Galactic Bulge}
The~abundance and distribution of planets in the Galactic bulge is still mostly unconstrained. Radial velocity and transit observations are biased toward nearby planetary systems and provide only sparse indication on bulge planets. Results from the SWEEPS 
 transiting survey~\cite{Sahu2006, Sahu2007} suggest that planets are as abundant in the Galactic bulge as they are in the solar neighborhood. However, the potentially large fraction of false positives prevents from drawing definite conclusions~\cite{Clarkson2008}. 

Microlensing events, although rare, can extend the search for planetary candidates to larger distances, since they do not depend on the flux of the host star. Microlensing provided the first possible evidence of a super-Earth in the Galactic bulge~\cite{Nagakane2017} and of the first possible planet in the habitable zone of a bulge star~\cite{Batista2013}. More generally, existing microlensing data suggest that there is at least one bound planet per star in the Milky Way, including the bulge~\cite{Cassan2012}. On the other hand, the comparison of microlensing events with models of the Galactic distribution seems to indicate a low (less than 50\%) relative abundance of planets in the bulge compared to that in the disk~\cite{Penny2016}, although a fraction closer to unity cannot be ruled out. Future observations from Euclid~\cite{Penny2013} and WFIRST  
\cite{Montet2017}, as well as analysis of present campaigns such as Spitzer~\cite{Zhu2017}, will refine the constraint on the abundance ratio of planets in the bulge relative to the disk. For~the present study, we will assume that such a ratio is not significantly lower than unity. 

It must be noticed that the bulge is expected to be an older component of the Galaxy, with a predominantly old stellar population~\cite{Gonzalez2015}. Observations towards low extinction regions of the bulge allow a look into the turn-off position of the color-magnitude diagram, thus giving a reliable indicator for the mean age of stars. A solid conclusion is that only a small fraction (3.5\%) of the bulge stars can be younger than 5 Gyr and that the mean age is $\sim$10 Gyr~\cite{Clarkson2008, Clarkson2011}. This implies that, in principle, most planets in the bulge may have existed billions of years before Earth, with plenty of time to evolve life and even multicellular organisms, if conditions were suitable.

\subsection{Metallicity}

A relevant factor in most studies of galactic habitability is the metallicity of the host stars. Theoretical models suggest that terrestrial planets form only above a threshold in metallicity of order $0.1$ of the solar value~\cite{Johnson2012}. Observations found small planets with equal frequency around stars in a range between 0.25 and 2.5 of solar metallicity~\cite{Buchhave2012,Buchhave2014,Buchhave2015}. On the other hand, Jupiter-like planets tend to form preferentially at higher metallicities, with a probability scaling roughly as $P_J\propto 10^{2 Z_g}$, where $Z_g$ is the gas-phase metallicity~\cite{Wang2015}. This has led to the suggestion that too low metallicities can hinder the formation of habitable rocky planets, whereas too high metallicity can result in overproduction of hot Jupiters, with the consequent disruption of inner orbits in the circumstellar habitable zone~\cite{Lineweaver2004}. Therefore, it is generally expected that habitable planets are found preferentially at intermediate metallicities, closer to the solar value. As in the disc the metallicity at any given epoch increases moving toward the center, this has led to regard the habitability of the inner Galaxy unfavorably, because of the  excessive abundance of heavier elements. 

However, there is evidence that the metallicity of the Galactic bulge behaves differently from that of the disc. With the development of multi-object spectrographs and more thorough observations in many fields of the bulge (not just the low-extinction regions), it has become apparent that the bulge is not simply made of old metal-rich stars. Observations of the bulge revealed a broad metallicity distribution~\cite{Zoccali2003} with $Z$ values ranging from $-$1.0 to 0.4, peaking at solar metallicity. Such a varying mean metallicity is unusual for just one stellar population. Instead, the idea of a bimodal stellar population in the bulge has gained consensus~\cite{Hill2011}. Recent Gaia-ESO 
observations of the bulge~\cite{Rojas-Arriagada2014} seem to support the conclusion of at least two metallicity components: a metal-poor and separately a metal-rich component as well~\cite{Gonzalez2015}. Higher metallicity stars tend to be concentrated toward the inner portions of the bulge along the plane, while metal poor stars seem to be predominantly further from the plane. 

These considerations provisionally suggest that metallicity in the bulge is not incompatible with the presence of rocky habitable planets. In itself, the abundance of heavier elements does not seem to be a showstopper to Galactic habitability even at distances smaller than 2.5 kpc from the center. 

\subsection{Orbital Stability and Encounters}

A separate question, more directly related to habitability, is whether planetary orbital configurations can remain stable for extended periods (comparable to those involved in biological evolution) in a dense stellar environment. Early studies suggest that habitable planets are unlikely in long-lived stellar clusters, as they can be expelled from their host system~\cite{DeJuan2012}. Others find that the rate of dangerous encounters (defined as gravitational interactions capable of moving the planet away from the circumstellar habitable zone) between Earth-like planets and stars is negligible, even in the Galactic bulge, being  $\sim$$7\times 10^{-4}$ Gyr$^{-1}$ for stellar densities $10^4$ times larger than in the solar neighborhood~\cite{Bojnordi-Arbab2020, Sloan2017}. A recent numerical analysis has estimated that 80\% of the stars in the bulge have at least one encounter below 1000 AU over a time of 1 Gyr~\cite{McTier2020}. Whether this would pose detrimental effects on the habitability of planets in the bulge would require further investigation.

A specific treatment has been reserved to the so called S-stars nearest to the Galactic center, with regard to their interaction with the supermassive black hole Sgr A*: even in such extreme case, it is not evident that the orbit of planets around such stars would be significantly affected~\cite{Trani2016, Davari2019a, Davari2019b, Iorio2020}. 

We can tentatively assume that stellar encounters, although more frequent, do not present special concerns in the bulge with respect to the disk. Anyway, we point out that assessing the habitability conditions in the bulge is not limited to planets in bound orbits around a host star: it would be relevant also for possible rogue, free-floating planets that may host life, for example below an icy surface~\cite{Abbot2011, Lingam2019b}.  

\subsection{Planets Around the Central Black Hole}

Stars are not the only astrophysical objects surrounded by disks of gas and dust, and they might not be the only sites of planet formation. Such disks are also often found surrounding supermassive black holes (SMBH). According to the unified model of AGNs 
\cite{Antonucci1993, Netzer1995}, gas and dust are found as a geometrically and optically thick torus, obscuring the broad emission line (width of several \mbox{1000 km\, s$^{-1}$}) region around the central accretion disk. It has been suggested that this disk can be the site of planetary formation~\cite{Wada2019} around low luminosity Seyfert-type AGNs rather than in quasar-type high luminosity ones with massive SMBHs. These planets would typically be of masses ten times larger than Earth. Thus, such exotic planets would resemble not only rocky worlds but could also be Neptune-like icy giant planets. Perhaps around $10^4$ planets of such kind could orbit each dim active galactic nucleus. This, however, would represent only a small fraction of the stellar density around the nucleus (which is about $10^7$ within a radius of 1 pc). 

The~center of the Milky Way hosts an SMBH (called Sgr A*), that is currently in a quiescent phase. Simulations show~\cite{Wada2019} that some planets might exist around it, orbiting 10 to 30 light-years from the nucleus, thus at a safe distance from the strong gravitational field and tidal forces. Standard exoplanet observing techniques are not applicable for hunting these worlds. Even X-ray interferometer in space might not be useful, because the occultation of the accretion disk by the planets would be indistinguishable from the intrinsic time variability of AGNs. An indirect detection method could rely on spectral changes at millimeter-wavelength due to opacity variation associated with dust growth, a technique often used when observing protoplanetary disks.
	
The~thermodynamics of exoplanets around SMBH, in principle, do not exclude the existence of life based on known biology. As shown in~\cite{Bakala2020}, hypothetical exoplanets at very low Keplerian circular orbits, in close vicinity of rapidly spinning supermassive black holes, could be kept far from equilibrium. Proposed mechanisms include the heat of strongly blueshifted and focused flux from the cosmic microwave background (CMB), and the cooling of the local sky containing the black hole shadow. This is similar to planets around stars heated by star radiation and cooled by the night sky. Thus, such temperature difference might, in principle, support processes far from thermodynamic equilibrium, including life, on exoplanets around SMBH. 

\section{The~Radiation Environment in the Bulge}
If rocky planets can exist around stars in the Galactic bulge, the dominant hindrance to habitability would be the presence of higher levels of ionizing radiation in the inner Galactic regions with respect to the outer disc. We then examine the consequences of the two main aspects that contribute to the radiation environment in the bulge: the presence of the SMBH in the Galactic center and the occurrence of transient radiation outpouring from supernovae (SN) explosions. While, in principle, gamma ray bursts (GRB) could also be included as a potential source of transient radiation events, their contribution is most relevant in the outskirts of the galaxy, as they are preferentially observed at low metallicities~\cite{Piran2014, Gowanlock2016}. Therefore, we neglect their effect in our analysis. 

\subsection{The~Supermassive Black Hole} 

Several studies have investigated the impact of the SMBH at the center of the Milky Way on Galactic habitability~\cite{Balbi2017, Forbes2018, Wislocka2019, Chen2018, Lingam2019a}. The~most relevant effect is the atmospheric loss of terrestrial planets due to the cumulative high-energy radiation released by the SMBH at the peak of its accretion phase. This~can be estimated through a simple energy-limited approximation to hydrodynamical escape, resulting in a total atmospheric mass lost for a planet at distance $D$ from the galactic center over a time $\Delta t$~\cite{Balbi2017}:
\begin{equation}
M_{\rm lost} = \frac{3\epsilon }{12\pi} \frac{L_{\rm XUV}\Delta t}{G\rho_p D^2}e^{-\tau}\end{equation}
where $L_{\rm XUV}$ is the SMBH luminosity in the X-ray and extreme ultraviolet (XUV), of order $10^{43}$ erg~s$^{-1}$, $\rho_p$ is the planet bulk density and $\epsilon$ is an efficiency parameter that is estimated in the range $0.1\leq \epsilon\leq 0.6$. The~exponential term accounts for the possible attenuation of the radiation due to the torus surrounding the SMBH, and depends on the XUV optical depth $\tau$. Depending on the choice of parameters, for planets with terrestrial density, a loss as large as the present-day Earth atmospheric mass can occur in a range of distances 0.2--1 kpc from the galactic center~\cite{Balbi2017}. A similar approach has been used to calculate the possible atmospheric mass loss for 54 known exoplanets in the Milky Way, of which 16 were hot Jupiters residing in the Galactic bulge, showing that they may have lost up to several Earth atmospheres in mass during the AGN phase of Sgr A*~\cite{Wislocka2019}. These results do not take into account the possible regeneration of atmosphere due to outgassing from the planetary interior, and they should be taken as an upper limit to the amount of evaporation. Furthermore, the dominant mechanisms for loss of secondary atmospheres is not hydrodynamic escape but rather non-thermal stellar-wind mediated escape (see, e.g.,~\cite{Lingam2019i}).

The~biological consequences of the irradiation from Sgr A* during its active phase are hard to assess, as they depend not just on the radiation tolerance of specific biota, but also on the amount of screening provided by the residual atmosphere. In the case of a totally exposed planetary surface, the level of ionizing radiation during the active phase of Sgr A* could prove lethal for unicellular radiation-resistant organisms up to 1 kpc from the Galactic center~\cite{Balbi2017}. Different estimates of the biological effect based on various assumptions for the atmospheric composition of Earth-like planets~\cite{Lingam2019a} showed that a distance of 1 kpc may, in fact, represent an upper limit to the harmful influence of the central AGN, and that a more likely value to produce significant damage on planets protected by an Earth-like atmosphere is of the order of only a few pc. 

While severe atmospheric loss can undoubtedly affect the habitability of Earth-like planets in the bulge, it must be kept in mind that the the active phase of the SMBH occurred during the initial stage of the Galaxy formation and it is expected to have lasted far less than a Gyr. Therefore the estimates above do not apply to planets that formed at later times. Moreover, as pointed out in~\cite{Chen2018}, rocky planets that start with a large gas envelope, such as super-Earths or mini-Neptunes, can instead become more habitable after experiencing substantial atmospheric loss. Similarly, the irradiation from high-energy flux can, in principle, stimulate biogenic effects~\cite{Lingam2019a}, proving beneficial rather than harmful for life.

A separate concern arising from the presence of a SMBH is related to effects that can last or repeat themselves over the entire galactic history, e.g., from charged particle winds or tidal disruption events. The~impact of such effects on habitability is still poorly understood and would certainly deserve further~investigation.

\subsection{Supernovae Explosions}
By far, the most important deterrent to the habitability of the inner region of the Galaxy emerging from previous studies is the highest density of stars, resulting in increasing risk of nearby supernovae (SN) explosions. 

The~most probable occurrence of SN is expected to be from Type II: these are the endpoint of stars with masses above $8 M_\odot$. While Type Ia SN can individually be more luminous and, therefore, more dangerous, they are also less frequent, as their progenitors are white dwarfs in binary systems, which is only $1\%$ of the stars in the range $0.08 M\odot < M < 8 M_\odot$~\cite{Gowanlock2011}.
The~distance at which a SN can significantly affect the habitability of an Earth-like planet is usually estimated at $R \le 8$ pc~\cite{Gehrels2003}. This~estimate is based on a 30\% depletion of atmospheric ozone from the SN blast, which results in the doubling of UV radiation from the host star. This enhanced level of ionizing radiation can trigger mass extinctions of complex land-based life~\cite{Melott2011}. However, the complete eradication of life from an Earth-like planet would require the full evaporation of oceans, as well as high-energy fluences well above the survivability threshold of radiation-tolerant extremophile organisms. This would likely require a much closer event, at distance $R \le 0.04$ pc~\cite{Sloan2017}. 

\begin{figure}[H]
\centering
\includegraphics[width=0.7\textwidth]{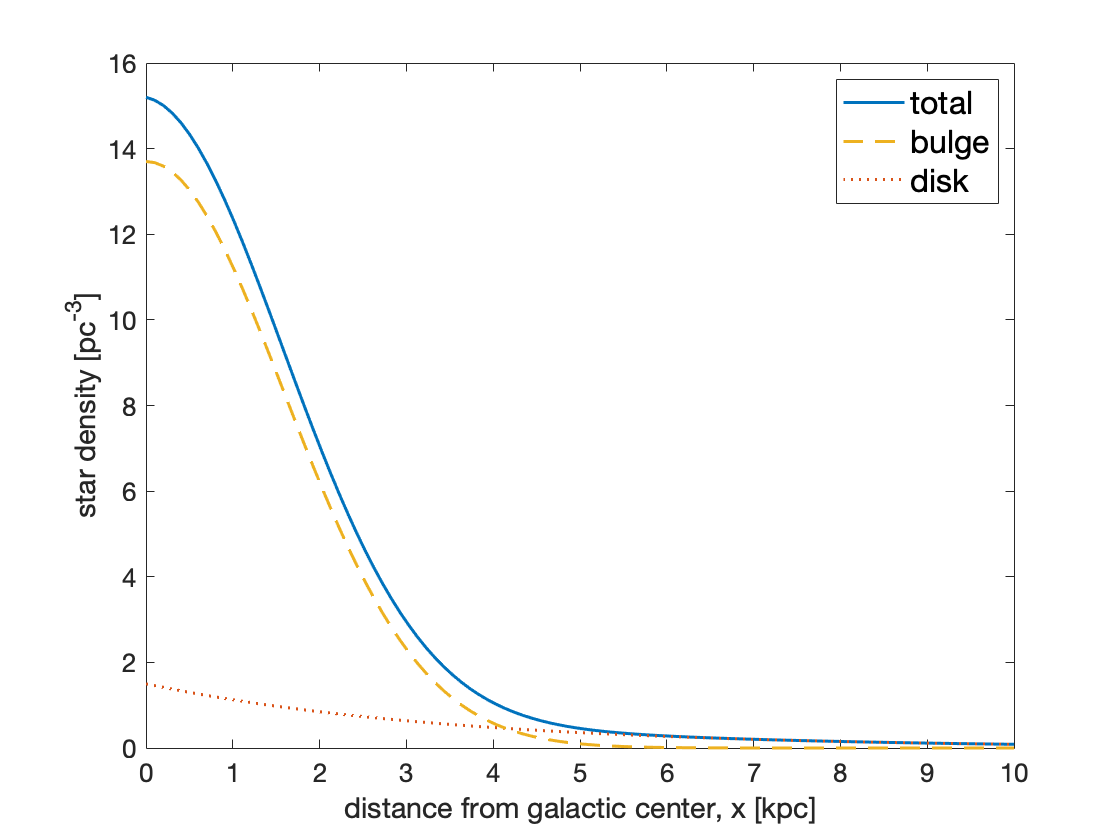}
\caption{Number density profile of stars in the galaxy along the $x$ axis (with $y=z=0$), according to the analytic model described in the text. \label{stardensity}}
\end{figure}

No detailed analysis of the impact of SN explosions on planets in the bulge has been performed so far. Some insight on the issue can be obtained by adopting an analytic model of the star density in the Galaxy and estimating the expected number of nearby SN at any given location. The~density of stars in the bulge can be modeled as a triaxial Gaussian profile~\cite{Robin2003}:
\begin{equation}
n_B(x,y,z)=n_{B0}\exp{(- r_s^2/2)}
\end{equation} 
for $\left(x^2+y^2\right)^{1/2}<2.5$ kpc, where $n_{B0}=13.7$ stars/pc$^3$ and \begin{equation}
r_s=\left(\left[\left(x/ x_0\right)^2+\left(y/y_0\right)^2\right]^2+\left(z/z_0\right)^4\right)^{1/4},
\end{equation} with $x_0=1.59$ kpc, $y_0=0.424$ kpc, and $z_0=0.424$ kpc. The~$x$ and $y$ axes lie in the galactic plane, with the $x$ axis forming an angle $30^\circ$ with the direction perpendicular to the line between the Sun and the Galactic center.
The~star density in the disk can be modeled as~\cite{Zhu2017}:
\begin{equation}
n_D(x,y,z)=n_{D0}\exp{\left[-\left(\frac{(r-r_{GC})}{r_0} +\frac{\vert z\vert}{z_{D0}}\right)\right]}
\end{equation} 
with $n_{D0}=0.14$ pc$^{-3}$, $r=\sqrt{x^2+y^2}$, $z_{D0}=325$ pc, $r_0=3.5$ kpc, and $r_{GC}=8.3$ kpc.
Thus, the total star density at any location (Figure~\ref{stardensity}) is given by:
\begin{equation}
n(x,y,z)=n_B(x,y,z)+n_D(x,y,z).
\end{equation}

At the Sun galactocentric distance ($x=8.3$ kpc, $y=z=0$), the average long-term SN rate within a spherical volume of radius $R$ smaller than 100 pc is estimated as~\cite{Melott2011}:
\begin{equation}
P_{\rm SN,\odot}(<R) = 2 \times 10^{-6} {\rm yr}^{-1} \left(\frac{R}{100{\rm\ pc}}\right)^3
\end{equation}

We can then estimate the SN rate within a spherical volume at any location by rescaling with the appropriate star density: 
\begin{equation}
P_{\rm SN}(<R) = P_{\rm SN,\odot}(<R) n(x,y,z) / n_\odot
\end{equation}
where $n_\odot=n(8.3 {\rm\ kpc}, 0, 0)$ is the star density at the Sun location. 

\begin{figure}[H]
\includegraphics[width=0.5\textwidth]{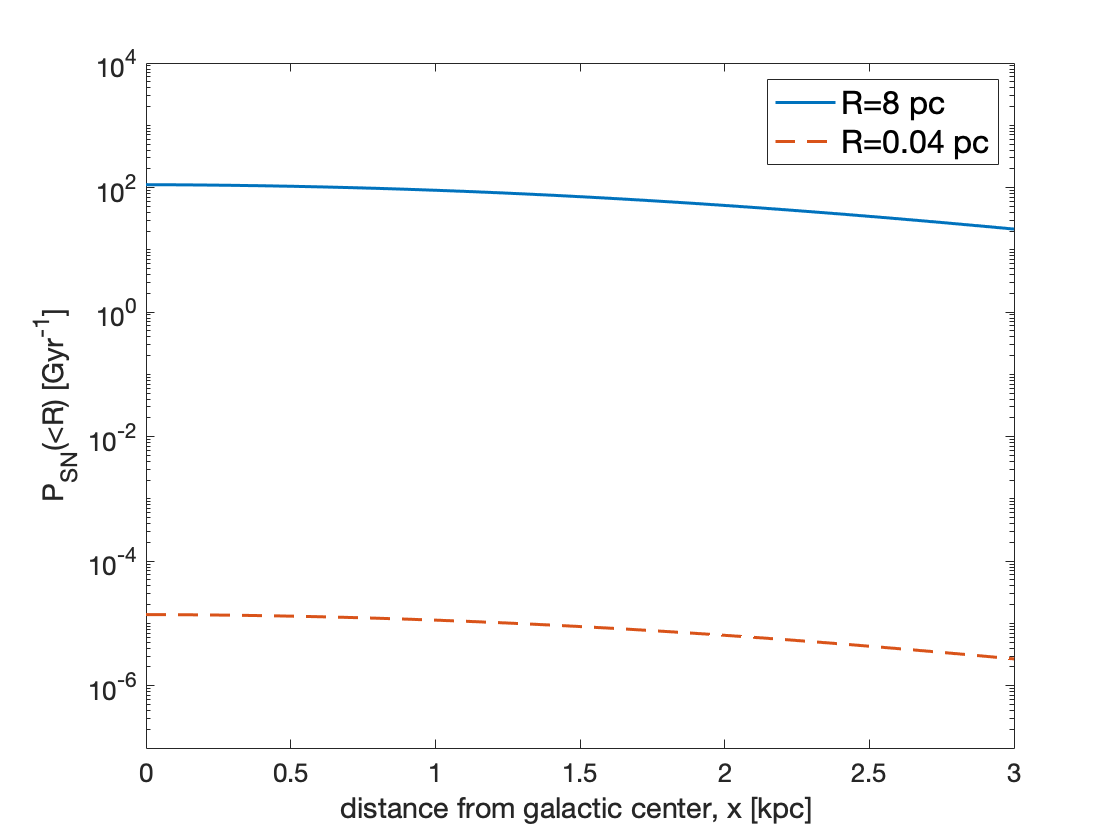}
\includegraphics[width=0.5\textwidth]{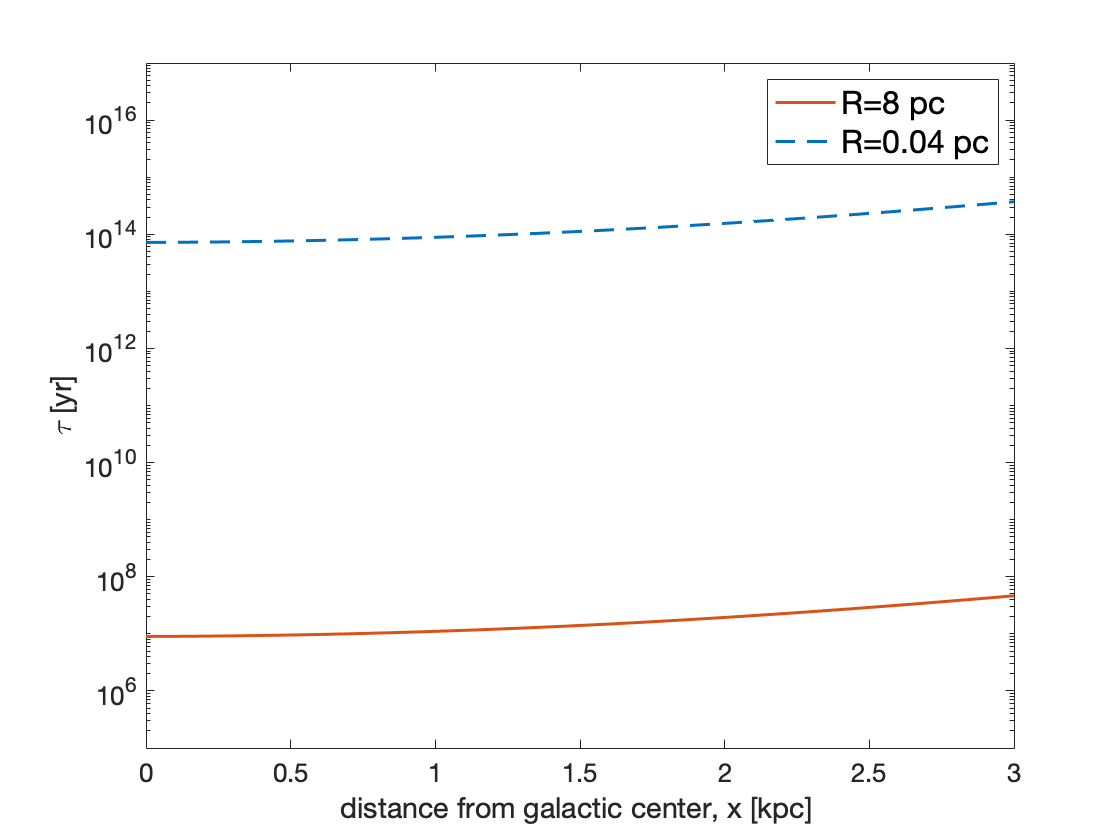}
\caption{The~expected rate of supernovae (SN) explosions (\textbf{left} panel) and the average interval between them (\textbf{right}~panel) as a function of the distance from the galactic center along the $x$ axis (with $y=z=0$). The~two lines are the rate of events occurring at a given location within a sphere of radius $R=8$ pc (continuous line) and $R=0.04$ pc (dashed line). These limits represent, respectively, the minimum distance for a mass extinction event and for a complete sterilization. \label{snrate}}
\end{figure}

This is shown in Figure~\ref{snrate} (left panel) for the two fiducial values of $R$ corresponding to mass extinction (8 pc) and complete sterilization (0.04 pc). It is apparent that, also at the relatively high stellar densities in the bulge, the probability that a nearby SN explosion completely sterilizes a planet is quite low: the expected number of such events per Gyr, at locations closer than 2 kpc from the Galactic center, is {$\sim$}$6\times 10^{-6}$--$1.4 \times 10^{-5}$. The~situation is different for SN explosions that, though not making a planet inhabitable, can cause significant harm to the atmosphere and to land life: planets in the bulge can expect between 40 and 110 extinction level events per Gyr. For~comparison, the expected current rate of such events in the solar neighborhood is {$\sim$}1 per Gyr~\cite{Melott2011}.

We can estimate the average interval between events as $\tau\sim P_{\rm SN}^{-1}(<R)$. This is shown in Figure~\ref{snrate} (right panel). The~time between mass extinction events is in the range 1--5~$\times 10^7$ years: interestingly, this is comparable to the mass extinction periodic timescale of 20--60 Myr on Earth, as well as what may arise from large superflares
\cite{Lingam2017f}. The~interval between complete sterilization events
exceeds the age of the Galaxy. 

A caveat worth mentioning is that we treated the stars as stationary, while their velocity dispersion in the bulge can be substantially higher than in the disk. We also point out that our model does not incorporate a self-consistent treatment of star formation in the bulge. In this respect, we note that if we assume that star formation for the bulge happened in a single burst, as in inside-out formation scenarios, then most stars that could potentially become SNII did it over the first few tens million years from the formation of the Milky Way. This would leave plenty of time for planets to evolve conditions suitable for life at later epochs. Thus, the results presented here should probably be seen as a worst-case scenario.

\section{Transfer of Biological Material}

\begin{figure}[H]
\includegraphics[width=0.5\textwidth]{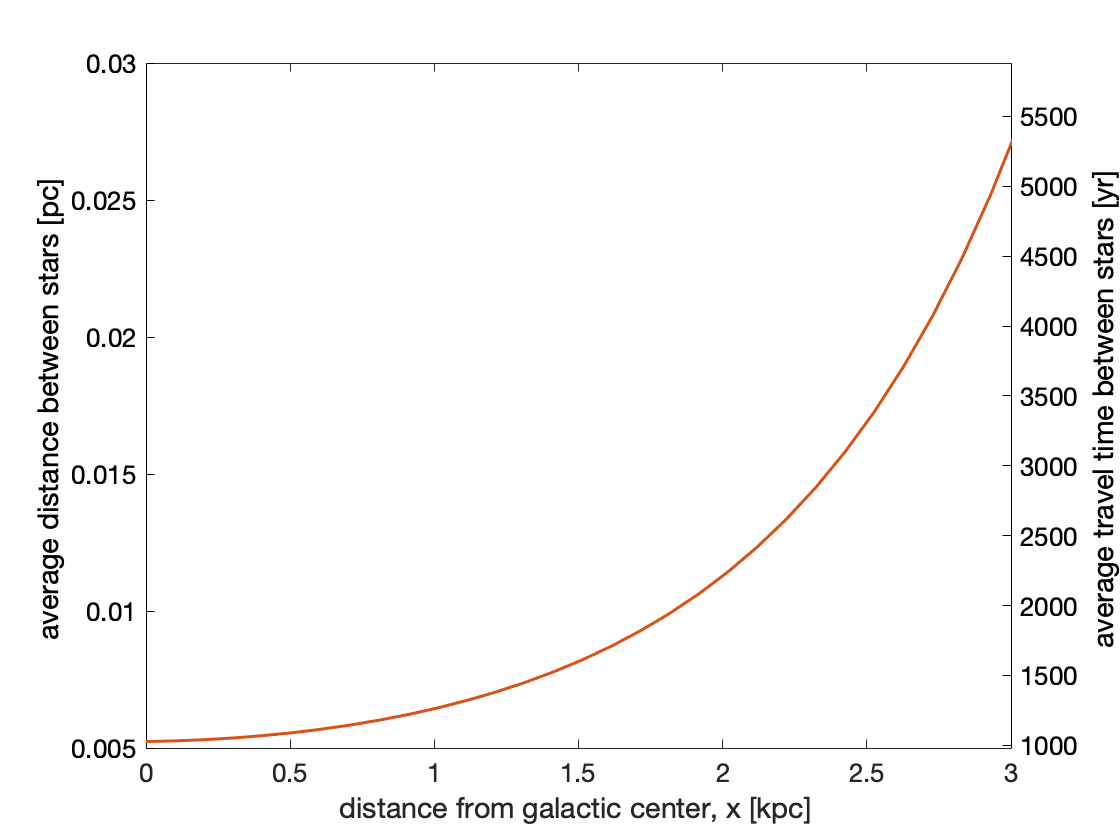}
\includegraphics[width=0.5\textwidth]{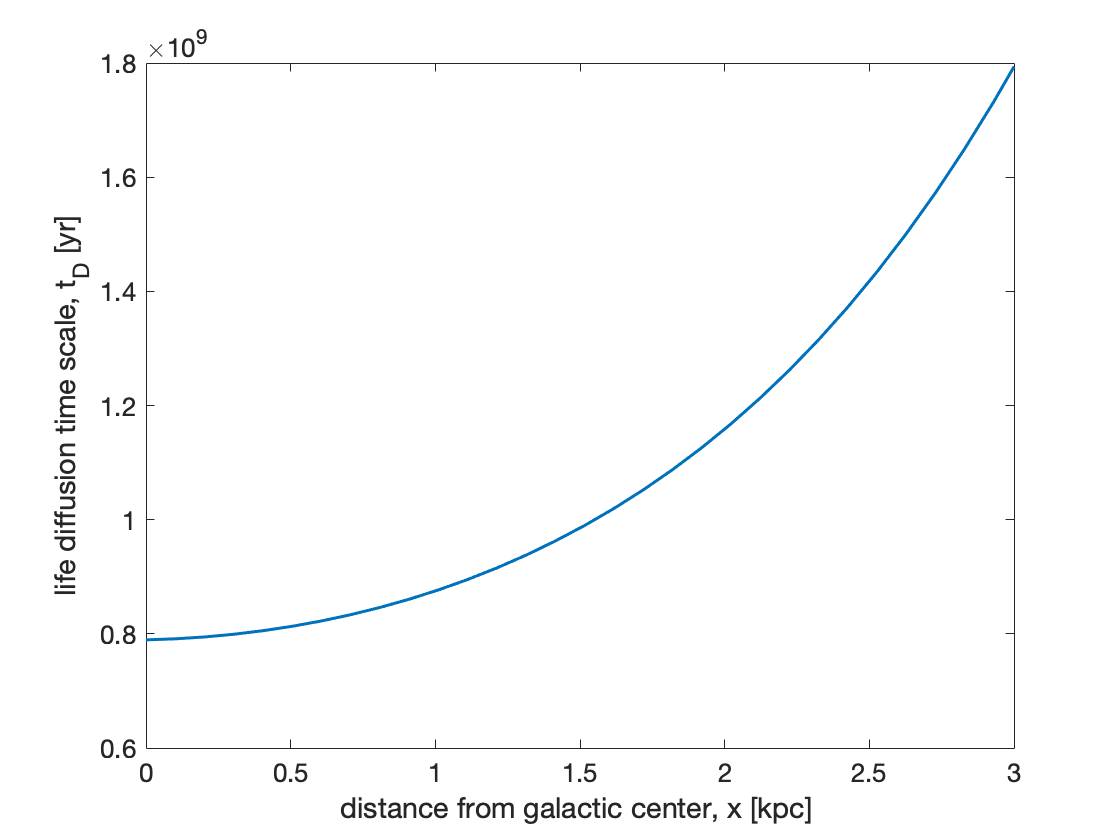}
\caption{The~average separation between stellar systems (\textbf{left} panel) and the time scale of life diffusion through lithopanspermia (\textbf{right} panel), as a function of the distance from the galactic center along the $x$ axis (with $y=z=0$). Shown on the right vertical axis of the left panel is the travel time for a given separation, assuming a velocity $v_e\sim 5$ km/s. \label{avdistance}}
\end{figure}

One potential advantage of the high density of stellar systems in the bulge might be the possibility of transferring active biological material over interstellar distances. Rocks containing microorganisms could be expelled from a life-bearing planet and be captured at later times by habitable planets elsewhere, possibly seeding the new location with life---a mechanism called `lithopanspermia'~\cite{Melosh1988, Wesson2010}. The~exchange of rocky material between stellar systems is by now observationally well-established~\cite{Meech2017}. The~survivability of microorganisms during the transport between planetary systems depends on the shielding mechanism provided by the rocks, which are required to have masses of $\sim$1--10~kg. The~fraction of surviving microorganisms after a travel time $t$ can be modeled as $P\propto \exp{(-t/\tau_s)}$, where $\tau_s$ is the typical survival time~\cite{Ginsburg2018}. There is no exact estimate for $\tau_s$, although values of order $\sim$$10^5$ years or higher seem possible given the right conditions. Figure~\ref{avdistance} (left panel) shows the average separation $\sim$$(3/4\pi n)^{1/3}$ between stellar systems in the bulge as a function of the distance from the galactic center. Assuming a plausible velocity for the ejecta, $v_e\sim$~5 km/s, the nearest neighboring systems can be reached in travel times of order a few thousand years, at least one order of magnitude smaller than the typical values in the galactic disk, and well within the estimated survival capability of terrestrial extremophiles. 

The~probability that an ejected rock can be captured by another terrestrial planet in a different stellar system has been first estimated for the solar neighborhood and deemed too small to be relevant as a life-spreading mechanism~\cite{Melosh2003}. However, it was subsequently shown that the capture probability could increase in crowded environments, such as in star-forming clusters~\cite{Adams2005, Belbruno2012} or in densely packed planetary systems~\cite{Lingam2017c}. Thus, it makes sense to ask whether lithopanspermia can play an essential role in the bulge compared to the disk. 

An initial estimate can be obtained by adopting the model outlined in~\cite{Melosh2003,Adams2005} for the rate of life-bearing rocks impacting a terrestrial planet in another stellar system:
\begin{equation}
\Gamma = \sigma v n_L
\end{equation}
where $v$ is the relative velocity of rocks with respect to the stars, $n_L$ is the number density of life-bearing rocks and $\sigma$ is the impact cross-section, the product of the capture cross-section from a stellar system, $\sigma_c$, and the probability that a rock  impacts a terrestrial planet in the system once is captured, $P_{\rm impact}$. The~capture cross-section is expected to vary based on the average stellar velocity dispersion, orbital configurations and multiplicity of the stellar and planetary systems, ejection velocity, rock size distribution, and so on. Plausible values are in the range $0.01-0.05$ AU$^2$~\cite{Melosh2003}. Here we have adopted the values $\sigma_c=0.025$ AU$^2$ and $P_{\rm impact}=10^{-4}$ from~\cite{Melosh2003}. We note that these are probably conservative in general, and in particular with respect to the conditions in the bulge. In fact, the value for $\sigma_c$ applies to planetary systems with a Jupiter-type planet in a Jupiter-like orbit. However, only $\sim$10\% of all systems meet this criterion. Binary star systems, that make up around $40\%$ of all stars, have a much higher cross-section~\cite{Lingam2018a,Ginsburg2018}. For~example, using the fit for $\sigma_c$ from~\cite{Adams2005} and assuming a velocity dispersion $\sim$120 km/s for stars in the bulge~\cite{Zhu2017} would result in a value $\sigma_c=0.045$ AU$^2$. 

The~number density of life-bearing rocks per year can be assumed to be proportional to the star density,  $n_L= \gamma n t$ with $\gamma\sim 15$/yr~\cite{Adams2005}. Then, the typical diffusion timescale for life between stellar systems can be found by $t=1/\Gamma$ and is:
\begin{equation}
t_{\rm D} = (\sigma v \gamma n)^{-1/2}
\end{equation}

This represents the evolution timescale of the fraction of inhabited planetary systems in the bulge, and is shown in Figure~\ref{avdistance} (right panel). All over the bulge, even a single inhabited planet could spread life to all other suitable stellar systems in a time $\sim$1 Gyr, much smaller than the age of the Galaxy.

\section{Discussion}

We have attempted a new assessment of the habitability of the inner 2 kpc regions (bulge) of the Milky Way. Although the actual frequency of terrestrial planets in the bulge is currently still mostly unknown, there is no compelling argument for ruling out the possibility that it is comparable to what is found in the disk region. 

The~main concern for the presence of suitable conditions for life in the bulge is the radiation environment from transient sources and the central SMBH. The~latter has been addressed in previous works and recognized as a strong source of ionizing radiation during its active phase, possibly leading to detrimental effects for habitability over the full extent of the bulge. In the present paper, we have evaluated the rate of SN explosions in the bulge using simple analytical models and found that, although much higher than in the disk, it is not likely to result in the complete eradication of life on inhabited planets. SN explosions, however, can cause mass extinction events with a frequency of $10^{-7}$/yr in the bulge, thus providing a difficult environment for life. The~actual impact of such occurrences for the habitability of planets in the bulge, however, would depend on many factors, including the initial atmospheric composition, the contribution from outgassing and the recovery capabilities of the biosphere. Moreover, the consequences would be different for land multicellular life---that may be more vulnerable to increased radiation levels and to severe disturbances to the atmospheric chemistry---and microorganisms adapted to a higher radiation environment. For~example, life was abundant on Earth much before the formation of an ozone layer, i.e., for $\sim$2~Gyr. Furthermore, life could survive underwater during extinction events and colonize the land in the intervals. For~comparison, after the Permian-Triassic event---the largest mass extinction recorded on Earth in the last $0.5$ Gyr---it took about 20 Myr for complete biotic recovery, and probably just a few million years for marine life~\cite{Brayard2017}.  It is then conceivable that a planet can remain habitable and host life even with a SN rate much higher than in the disk. 

We have also shown that lithopanspermia scenarios would be more efficient in the bulge with respect to the solar neighborhood. If indeed the spreading of life between stellar systems is possible, then the highest rate of catastrophic events might be balanced by the possibility that life can migrate quickly to safer locations. Any attempt to model the possible distribution of inhabited planets in the inner region of the galaxy should therefore take this aspect into account. This would provide an additional motivation for a more thorough investigation of the bulge habitability. 

An interesting aspect that we did not address in this paper would be to establish a measure for the overall propensity to life in the bulge, based on detailed Monte Carlo simulations of the habitability/sterilization of stellar systems, combined with models of life appearance and recovery. Intuitively, one can foresee a trade-off between the higher risk of catastrophic events in the bulge and the greater abundance of stellar systems---with the latter increasing the potential number of locations that can host life. As shown in~\cite{Gowanlock2011,Morrison2015}, there is a definite increase in the propensity to life---measured as the existence of ``windows of opportunities'' between extinctions events---as one moves towards the galactic center. However, it is plausible that this trend may stop somewhere within the bulge, when the advantages of a highest density of stars are outweighed by the highest catastrophic risk. We~plan to investigate this point in future work. 

\vspace{6pt} 

\authorcontributions{Conceptualization, A.B. and A.K.; methodology, A.B. and A.K.; software, A.B., M.H., and A.K.; writing---original draft preparation, A.B.; writing---review and editing, A.B., M.H., and A.K.; supervision, A.B. and A.K. All authors have read and agreed to the published version of the manuscript.}

\funding{A.B. was partially funded by the Italian Space Agency through the Life in Space project (ASI N. 2019-3-U.0) and by grant number FQXi-MGA-1801 and FQXi-MGB-1924 from the Foundational Questions Institute and Fetzer Franklin Fund, a donor advised fund of Silicon Valley Community Foundation. A.K. was supported in this work by Ministry of Education, Science and Technological development of Republic Serbia through the  \emph{Astrophysical Spectroscopy of Extragalactic Objects} project number 176001. This research was supported by the Erasmus Mundus Master Program, AstroMundus.}

\acknowledgments{The~authors wish to thank Giuseppe Bono and Manasvi Lingam for useful discussions.}

\conflictsofinterest{The~authors declare no conflict of interest.} 

\abbreviations{The~following abbreviations are used in this manuscript:\\

\noindent 
\begin{tabular}{@{}ll}
AGN & Active Galactic Nucleus\\
GHZ & Galactic Habitable Zone\\
ESO & European Southern Observatory\\
SMBH & Supermassive black hole\\
SN & Supernova(e)\\
SWEEPS & Sagittarius Window Eclipsing Extrasolar Planet Search\\
WFIRST & Wide Field Infrared Survey Telescope (now Nancy Grace Roman Space Telescope)
\end{tabular}}

\reftitle{References}




\end{document}